 \documentclass[final,3p,times]{elsarticle}
\usepackage{lscape,epsfig}
\usepackage{amsfonts}
\usepackage{amsmath}
\usepackage{amssymb}
\usepackage{slashed}

\usepackage{graphicx}
\usepackage{epstopdf}
\usepackage{color}

\usepackage{ulem}

\usepackage{float}

\newcommand{\sign}{\text{sign}}
\DeclareMathOperator{\tr}{tr}

\def\beq{\begin{equation}}
\def\eeq{\end{equation}}
\def\beqa{\begin{eqnarray}}
\def\eeqa{\end{eqnarray}}
\def\ban{\begin{eqnarray*}}
\def\ean{\end{eqnarray*}}
\def\bi{\begin{itemize}}
\def\ei{\end{itemize}}

\def\ni{\noindent}

\journal{Nuclear Physics B}

\begin{document}

\begin{frontmatter}



\title{Magnetized pole-mass of neutral $\rho$ meson within full RPA evaluation}


\author[label1]{Sidney S. Avancini}
\ead{sidney.avancini@ufsc.br}

\affiliation[label1]{organization={Departamento de Física, Universidade Federal de Santa Catarina},%Department and rganization
 %           addressline={}, 
            city={Florianópolis},
            postcode={88040-900}, 
            state={SC},
            country={Brazil}}
            
\author[label2]{Ricardo . L. S. Farias}
\ead{ricardo.farias@ufsm.br}

\affiliation[label2]{organization={Departamento de Física, Universidade Federal de Santa Maria},%Department and rganization
            city={Santa Maria},
            postcode={97105-900}, 
            state={RS},
            country={Brazil}}

\author[label1]{William R. Tavares}
\ead{william.tavares@posgrad.ufsc.br}

\author[label3]{Varese S. Timóteo}
\ead{varese@ft.unicamp.br}

\affiliation[label3]{organization={Grupo de \'Optica e Modelagem Num\'erica - GOMNI, Faculdade de Tecnologia - FT, 
Universidade Estadual de Campinas - UNICAMP},%Department and rganization
 %           addressline={}, 
            city={Limeira},
            postcode={13484-332}, 
            state={SP},
            country={Brazil}}

\begin{abstract}
In this work we calculate the pole-mass of the   $\rho^0$ meson  with different spin projections $s_z=0,\pm 1$ in the context of the
magnetized two-flavor Nambu--Jona--Lasinio model. Making use of the mean field approximation to obtain the effective quark mass as a 
function of the magnetic field, we apply the random phase approximation (RPA) to the vector channel in order to calculate the polarization
function for each spin component. We adopt the magnetic field independent regularization (MFIR) in our evaluations as a method of 
separating divergences and the Pauli-Villars regularization for the vacuum contributions. The $\rho^0$ meson mass with spin projection
$s_z=\pm 1$ is always catalysed with the magnitude of the magnetic field, showing good agreement with Lattice QCD results. The mass 
projection $s_z=0$ has a non-monotonic behavior, decreasing until the minimum at $eB\lesssim 0.15$ GeV$^2$, which is in contrast with 
available LQCD data. 

\end{abstract}



%\begin{keyword}



%\end{keyword}

\end{frontmatter}


\section{Introduction}
\label{Introdu}
The properties of hadronic matter under strong magnetic fields has been one of the most intense subject of research in the last few years.
Several phenomena can be explored in this context, as the magnetic catalysis \cite{Shovkovy:2012zn,Miransky:2015ava}, the chiral magnetic
effect \cite{Fukushima:2008xe} and the chiral separation effect \cite{Huang:2015oca}. From the experimental point of view, the measurements
can be explored in several types of non-central ultrarelativistic heavy ion collisions, that can reach magnetic fields of the order $eB\sim 10^{19}$
G \cite{Fukushima:2008xe}. Neutron stars presenting strong magnetic fields, also known as magnetars, are also a frequent object of study  
and the magnetic field of such objects can be of the order $eB\sim 10^{18}$ G \cite{Duncan:1992hi,Kouveliotou:1998ze}. It is also possible that
strong fields were present in the primordial universe \cite{Vachaspati:1991nm}.

Until now, one way to test some of the inaccessible possibilities of the experiments or the theory of Quantum Chromodynamics 
(as we are in the low energy regime) in strong magnetic fields is to use Lattice QCD (LQCD) or effective models like the NJL model \cite{Nambu12,Menezes:2008qt,FariasPRC,Farias:2016gmy,PhysRevD.94.054019,PhysRevD.96.034007,Ferrer:2013noa}, 
PNJL \cite{Ratti:2005jh,Fukushima:2010fe,Ferreira:2013oda}, Abelian Higgs Model~\cite{PhysRevD.88.036010,PhysRevD.89.116017}, Linear Sigma Model with quarks~\cite{Scavenius:2000qd,Fraga:2008qn,Schaefer:2009ui,Ayala:2021nhx}, 
Chiral Perturbation Theory \cite{Andersen:2012zc,Andersen:2014xxa}, Holographic prescriptions \cite{Callebaut:2011ab,Callebaut:2013wba} and approaches
based on the Schwinger-Dyson equations (DSE) \cite{dse}.  Some of the main results of these models are in accordance with the LQCD, 
as the magnetic catalysis of the chiral condensate at $T\sim 0$. On the other hand, an inverse magnetic catalysis (IMC) was predicted by 
LQCD \cite{Bali:2012zg} at temperatures close to the pseudocritical temperature. More details about MC and IMC can be seen in
\cite{imcreview}.

A very good way to test many properties of the quark matter under strong magnetic fields is to understand what is the behavior of the lightest
mesons and its properties. Recent lattice QCD
\cite{Bali:2015vua,Bali:2017ian,Luschevskaya:2014lga,lush1,Luschevskaya:2013fua,Luschevskaya:2012xd,Hidaka:2012mz,Ding:2022tqn} and several 
effective model studies have been focused on the calculation on the masses of neutral and charged pions  
\cite{Andersen:2012zc,Avancini:2016fgq,Avancini:2015ady,Avancini:2018svs,Fayazbakhsh:2012vr,Fayazbakhsh:2013cha,Mao:2017wmq,Zhang:2016qrl,GomezDumm:2017jij,Wang:2017vtn,Liu:2018zag,Mao:2018dqe,Ayala:2018zat,Coppola:2019uyr,Coppola:2018vkw,Ayala:2020muk,Klevansky:1991ey,Avancini:2021pmi,Sheng:2021evj}, $\rho^0$ and $\rho^{\pm}$ \cite{Andreichikov:2016ayj,Liu:2014uwa,Andreichikov:2013zba,Ghosh:2020qvg} an its properties. 
These results can be useful for extensions of the effetive models for application in beyond-mean-field approaches \cite{Mao:2016fha}. 
A particular situation is the magnetic inhibition \cite{Fukushima:2012kc}, where the excitation of $\pi^0$ at the pseudocritical temperature can
lead to a manifestation to the inverse magnetic catalysis. Also, the $\rho^{\pm}$ meson condensation 
\cite{Chernodub:2010qx,Chernodub:2011mc,Chernodub:2012zx,Li:2013aa} has been one of the most
attracting subjects for both effective models \cite{Andreichikov:2016ayj,Liu:2014uwa,Andreichikov:2013zba,Frasca:2013kka} and lattice QCD
calculations  \cite{Bali:2017ian,lush1,Hidaka:2012mz,Braguta:2011hq}. Although the LQCD is a very robust technique for the energy regime
that interests us, there are at some point limitations, as the sign problem \cite{muroya,Karsch}. A very useful approach for explore QCD 
at such conditions is to use effective models, like the NJL model, that enables us to explore the full QCD phase diagram in a 
more accessible way.

While handling the non-renormalizable four-point interaction models, it is possible to obtain spurious results associated with the regularization 
procedure with the model under strong magnetic fields \cite{Avancini:2019wed,Allen:2015paa}. These results can be outlined by separating 
properly the ultraviolet divergences from the medium part \cite{Schwinger:1951nm,Ebert:1999ht,Ebert:2003yk}. 
This procedure, known as MFIR, shows remarkable agreement with LQCD \cite{Avancini:2016fgq,Coppola:2018vkw,Avancini:2019wed}. 
Therefore, some works adopt the MFIR scheme in the NJL model to obtain, for example, the thermodynamical properties for quark matter 
by computing quantities like pressure, entropy, heat capacity, sound velocity and magnetization, amongst other quantities 
\cite{Menezes:2008qt,Farias:2016gmy,Ferreira:2013oda,Menezes:2009uc,Avancini:2012ee,Avancini:2020xqe,Ferreira:2014exa}. 
The MFIR method is also very useful to calculate the mesonic masses and its properties. 

In this work we investigate the neutral $\rho$ meson mass with different spin projections $s_z=0,1,-1$ (and without any polarization)  
in the $N_f=$2 NJL model using the MFIR prescription. We compared our results with the available lattice QCD results. The dependence of the spin 
projections on the free parameter, the vector coupling, is studied in detail in order to observe the regions where we obtain bound state solutions 
as function of the magnetic field strength. In the section \ref{formal0} we show all the analytical details of the $N_f=$2 NJL in a constant external magnetic field, 
as well as the development of the tensor functions and the meson mass with different projections. In section \ref{results} we show our numerical results and 
in the section \ref{conclusions} is devoted to our final remarks and main conclusions.

%%%%%%%%%%%%%%%%%%%%%%%%%%%%%%%%%%%%%%%%%%%%%%%%%
\section{General formalism}
\label{formal0}

The Lagrangian of the two-flavor ($N_f=2$) NJL model including a vector interaction in the presence of an external electromagnetic 
field is given by the following expression

\begin{eqnarray}
\mathcal{L}=\overline{\psi}\left(i \slashed D - \hat{m}\right)\psi
+G\left[(\overline{\psi}\psi)^{2}+(\overline{\psi}i\gamma_{5}\vec{\tau}\psi)^{2}\right]
-G_v\left[(\overline{\psi}\gamma^{\mu}\vec{\tau}\psi)^{2}+(\overline{\psi}\gamma^{\mu}\gamma_{5}\vec{\tau}\psi)^{2}\right]-
\frac{1}{4}F^{\mu\nu}F_{\mu\nu} ~,
\end{eqnarray}
where $A^\mu$, $F^{\mu\nu} = \partial^\mu A^\nu - \partial^\nu A^\mu$ 
are respectively the electromagnetic gauge and  tensor fields, $G$ and $G_v$ represents the scalar and vector coupling 
constants respectively, $\vec{\tau}$ are isospin Pauli matrices,  $Q$ is the diagonal quark charge matrix, 
$Q =  {\rm diag}(q_u, ~ q_d)$ = ${\rm diag}(2e/3, ~ -e/3)$,
$D^\mu =(\partial^{\mu}+iQA^{\mu})$ is the covariant derivative,   
 $\psi$ is the quark fermion field, and $\hat{m}$ represents the bare quark mass matrix\footnote{Our results are expressed in Gaussian natural units 
where $1\,{\rm GeV}^2= 1.44 \times 10^{19} \, G$, $\alpha=e^2/(4\pi)$, $e=1/\sqrt{137}$.}:
\begin{equation}
 \psi = \left(
\begin{array}{c}
\psi_u  \\
\psi_d \\
\end{array} \right) ~, ~
 \hat{m}  = \left(
\begin{array}{cc}
m_u & 0 \\
0 & m_d \\
\end{array} \right) ~.
\end{equation}
We consider here $m_u$=$m_d=m_0$  and choose the Landau gauge, $A^{\mu}=\delta_{\mu 2}x_{1}B$, which satisfies 
$\nabla \cdot \vec{A}=0$ and $\nabla \times \vec{A}=\vec{B}=B{\hat{e_{3}}}$, i. e., resulting in a constant magnetic field in the z-direction and
throughout this work the Minkowski metric $g^{\mu \nu} = {\rm diag} (1,-1,-1,-1)$ is assumed. In the mean field approximation, the Lagrangian 
$\mathcal{L}$ is denoted by

\begin{equation}
 \mathcal{L}=\overline{\psi}\left(i\slashed D-M\right)\psi- G \left \langle \overline{\psi}\psi \right \rangle^{2}-
 \frac{1}{4}F^{\mu\nu}F_{\mu\nu}~,
\end{equation}
\noindent where the constituent (effective) quark mass, $M$, is the solution of the gap equation 
\begin{equation}
 M = m_0-2G \left \langle \overline{\psi}\psi \right \rangle \; , 
\label{gap}
\end{equation}
and the quark condensate is given by 

\begin{eqnarray}
\left \langle \overline{\psi}\psi \right \rangle=-i  \int\frac{d^{4}k}{(2\pi)^4} \tr  \tilde{S}(k) \; , 
\label{cond}
\end{eqnarray}
where the trace has to be evaluated in color, flavor and Dirac spaces. The mean field quark propagator in coordinate space is given by 
$i\widetilde{S}(x,y)$=diag($i\widetilde{S}_u(x,y)$,$i\widetilde{S}_d(x,y)$) where 
\begin{equation}
 i\tilde{S}_f(x,y)= e^{i \Phi_f(x,y)} \; i\widetilde{S}_f (x-y)  \; , \; f=(u,d) \; .
\end{equation}
\noindent In the latter expression the propagator  is written in the Schwinger formalism \cite{Gusynin:1995nb} as a 
product of two terms, the Schwinger phase   
\begin{equation}
 \Phi_f(x,y) =\frac{iq_f}{2}(x-y)^{\mu}A_\mu(x+y) \; ,
\end{equation}
\noindent and a translationally invariant part 
\begin{eqnarray}
i\widetilde{S}_f (x-y) &=&  \int \frac{d^4 k}{(2\pi)^4} e^{-i k\cdot(x-y)}\; i \tilde{S}_f(k), \nonumber\\
 i \tilde{S}_f(k)&=&\int_0^\infty ds\exp\left[is\left(k_{\parallel}^2-k_{\perp}^2\frac{\tan(\beta_fs)}{\beta_fs}-M^2 + i\epsilon \right)\right]
 \times \left[(k_{\parallel}.\gamma_{\parallel}+M)\; \Pi_f(s)-k_{\perp} . \gamma_{\perp}\; g_f(s)\right] \;,\label{schwinger}
\end{eqnarray}

\noindent where $\beta_f=|q_fB|$, ${\Pi_f(s)=1-s_f\gamma^1\gamma^2\tan(\beta_fs)}$, ${s_f=\sign(q_f B)}$,  $a_{\parallel}.b_{\parallel}=a^0b^0-a^3b^3$, $a_{\perp}.b_{\perp}=a^1 b^1+a^2b^2$
and the function $g_f(s)=1+\tan^2(\beta_fs)$.

The gap equation in the MFIR scheme is given by the following expression \cite{Avancini:2019wed}
\begin{eqnarray}
\frac{M-m_0}{2G}-2MN_cI_1 -\frac{M^3N_c}{4\pi^2}\sum_{f=u,d}\eta(x_f)=0
\end{eqnarray}
where $N_c=3$ and $N_f=2$ are  the color and flavor numbers, respectively, and the finite magnetic contribution is given by 
\begin{eqnarray}
\eta(x_f)= \left[\frac{\ln\Gamma (x_f)}{x_f}-\frac{1}{2x_f}\ln 2\pi+ 1-\left(1-\frac{1}{2x_f}\right)\ln x_f\right],\nonumber\\
\end{eqnarray}
where $x_f=\frac{M^2}{2|q_fB|}$. The usual divergent non-magnetic contribution is given by $I_1$, which  may be written as  
\begin{equation}
 I_1=N_f \int \frac{d^3 k}{(2\pi)^3} \; \frac{1}{\sqrt{k^2+M^2}} \;  .
\end{equation}
A power counting shows that $I_1$ is ultraviolet divergent and needs to be regularized. The model properties depends on the regularization
prescription \cite{Klevansky:1992qe} and the latter has to be considered as part of the model definition. Here we adopt the Pauli-Villars 
regularization \cite{Klevansky:1992qe,Itzy} which preserves gauge invariance and has been shown to be adequate in the context of the NJL 
in the presence of a magnetic field \cite{Avancini:2019wed}.  

The  Pauli-Villars regularization consists of making the following substitution in the integrand:
\begin{align}
 &\frac{1}{\sqrt{k^2+M^2}} \to \sum_{i} c_i 
 \frac{1}{\sqrt{k^2+M_i^2}}  \; , \; 
   M_i^2 = M^2+a_i \Lambda^2 \; ,
\end{align}
\ni with the constraints
\begin{align}
   & \sum_i c_i =0 \quad , \quad \sum_i c_i \; M_i^2 = 0 \; .
\end{align}
The latter constraints assure that the integral $I_1$  is finite.   
Here we use two regulators, and the constants $c_0=1$ , $c_1=-2$ , $c_2=1$, $a_0=0$, 
$a_1=1$ and $a_2=2$ and $\Lambda$ is a regulating mass.  
Notice that the choice of the constants $c_i$ and $a_i$ is not unique and the one 
adopted here is often used in the literature \cite{Klevansky:1992qe,Itzy,Oertel:2000jp}.   
One can easily show that the regularized form of $I_1$ takes the form   
\begin{align}
I_1^{PV}=&\frac{\Lambda^2}{4\pi^2}\left[(2+M_{\Lambda}^2)\log(1+2M_{\Lambda}^{-2}) 
 -2(1+M_{\Lambda}^2)\log(1+M_{\Lambda}^{-2})\right],
\end{align}
where $M_{\Lambda}=\frac{M}{\Lambda}$. The numerical results using the Pauli-Villars regularization will be given in the Section \ref{formal0}.
%
% ----------------------------------------------------------------
%
\subsection{RPA evaluation of vector channel}

In the RPA formulation, the polarization tensors for each channel, i.e, pseudo-scalar, scalar and vector channels, are 
defined in coordinate space by

{\small \begin{align}
&\frac{1}{i}\Pi_{\pi}(x,y)=-  
\tr_{f,c,D}[i\gamma^{5}\tau^a i \tilde{S} (x,y)i\gamma^{5}\tau^b  
i \tilde{S} (y,x)],  \\
&\frac{1}{i}\Pi_{\sigma}(x,y)=- \tr_{f,c,D}[i \tilde{S} (x,y)  i\tilde{S} (y,x)],  \\
&\frac{1}{i}\Pi^{\mu\nu,ab}_{\rho}(x,y)=- \tr_{f,c,D}[\gamma^{\mu}\tau^{a} i\tilde{S} (x,y)\gamma^{\nu}\tau^{b}  i \tilde{S} (y,x)] \; .  
\end{align}} 

In this work, we consider only charged-neutral mesons since the Schwinger phase cancels out and it is possible to perform a standard Fourier 
transform in the polarization tensors, obtaining in momentum space:
{\small \begin{align}
&\frac{1}{i}\Pi_{\pi^0}(q^{2})=- \int \frac{d^{4}k}{(2\pi)^{4}} 
\tr_{f,c,D}[\gamma^{5}\tau^3\tilde{S} (k)\gamma^{5}\tau^3  \tilde{S} (k+q)],  \label{loop-ps}\\
&\frac{1}{i}\Pi_{\sigma}(q^{2})= \int \frac{d^{4}k}{(2\pi)^{4}}\tr_{f,c,D}[\tilde{S} (k)  \tilde{S} (k+q)],  \label{loop-s}\\
&\frac{1}{i}\Pi^{\mu\nu,33}_{\rho}(q^{2})= \int \frac{d^{4}k}{(2\pi)^{4}}\tr_{f,c,D}[\gamma^{\mu}\tau^{3}\tilde{S} (k)\gamma^{\nu}\tau^{3}  \tilde{S} (k+q)],  \label{loop-vector}
\end{align}}

\noindent where, the trace is evaluated in flavor, color and Dirac spaces. We use the definition of the polarization tensor $\Pi_{\rho}^{\mu\nu,ab}$ 
as given in Ref. \cite{Oertel:2000jp}. The general structure of the propagator in the vector channel, is given by the Schwinger-Dyson
equation \cite{Liu:2014uwa,Oertel:2000jp} as

\begin{eqnarray}
 D^{\mu\nu}_{ab}(q^2)=-2G_v\delta_{ab}g^{\mu\nu}+(2G_v\delta_{ac}g^{\mu\lambda})(\Pi_{\lambda\sigma,cd})(D^{\sigma\nu}_{db}). \nonumber\\
\end{eqnarray}

As we are interested only in neutral vector mesons,  
we define $\Pi^{\mu\nu}_{\rho} \equiv \Pi^{\mu\nu, 33}_{\rho}$ and 
$D^{\mu\nu} \equiv D^{\mu\nu, 33}$.
Thus, we obtain for the neutral $\rho$ meson the following simplified Schwinger-Dyson equation:
\begin{eqnarray}
 D^{\mu\nu}(q^2)=-2G_v g^{\mu\nu} + 2G_v g^{\mu\lambda}\; \Pi_{\rho \; \lambda\sigma} 
 \; D^{\sigma\nu} (q^2) \;,\label{propagator_rho0}
\end{eqnarray}
 and the polarization tensor also can be simplified further as
\begin{equation}
\frac{1}{i}\Pi^{\mu\nu}_{\rho}(q^{2})= N_c \sum_{f=u,d} \int \frac{d^{4}k}{(2\pi)^{4}}\tr_{D}[\gamma^{\mu}\tilde{S}_f (k)\gamma^{\nu} \tilde{S}_f (k+q)] \; .
\label{pol1}
\end{equation}

\ni Next, we consider the polarization tensor in the $\rho$ rest frame, i.e., 
$q^\mu\equiv(q^0=m_\rho,\vec{q}=\vec{0})$. From the explicit calculation of 
$\Pi^{\mu\nu}_{\rho}(q^{2})$ given in Eq. (\ref{pol1}) one obtains that
the only non-null components of the polarization tensor in its rest frame are
$\Pi^{1 1}_{\rho}(q^2_0)=\Pi^{2 2}_{\rho}(q^2_0)$ and $\Pi^{3 3}_{\rho}(q^2_0)$. 
Notice that the component $\Pi^{00}_{\rho}(q_0^2)$ is equal to zero due to the Ward identity, $q_{\mu}\Pi^{\mu\nu}_{\rho}(q_0^2)=0$.
In the following sections explicit expressions for these components will be given.
The projection structure of the polarization function can be made explicit by defining  
the following ortho-normal basis \cite{Liu:2014uwa} 
\begin{eqnarray}
&& \epsilon_1^{\mu}=\frac{1}{\sqrt{2}}(0,1,i,0), \nonumber \\
&& \epsilon_2^{\mu}=\frac{1}{\sqrt{2}}(0,1,-i,0), \nonumber \\
&& b^{\mu}=(0,0,0,1),  \nonumber\\
&& u^{\mu}=(1,0,0,0).
\end{eqnarray}
Thus the the polarization function can be written as
\begin{align}
\Pi^{\mu\nu}_{\rho}=\Pi^{11}_{\rho}\epsilon^{\mu}_1\epsilon^{*\nu}_1+ \Pi^{22}_{\rho}\epsilon^{\mu}_2 \epsilon^{*\nu}_2 +\Pi^{33}_{\rho}b^{\mu}b^{\nu}, \label{proj} 
\end{align}
where $\epsilon^{\mu}_1\epsilon^{*\nu}_1$, $\epsilon^{\mu}_2\epsilon^{*\nu}_2$ 
and $b^{\mu}b^{\nu}$ are associated respectively to the $s_z=1$, $s_z=-1$ and $s_z=0$ components  of the spin projectors.

Now, using the Schwinger-Dyson equation Eq. (\ref{propagator_rho0}) and the previous structure of the polarization function 
the transverse part of the generalized rho meson propagator in the RPA formulation reads:
\begin{equation}
D^{\mu\nu}=D^{11}\epsilon^{\mu}_1\epsilon^{*\nu}_1+D^{22}\epsilon^{\mu}_2\epsilon^{*\nu}_2+D^{33}b^{\mu}b^{\nu}, 
\end{equation}
where the $D^{ii}$  non-null components are given by
\begin{eqnarray}
&& D^{11}(q_0^2)=\frac{-2G_v}{1-2G_v\Pi^{11}_{\rho}(q_0^2)},\\
&& D^{22}(q_0^2)=\frac{-2G_v}{1-2G_v\Pi^{22}_{\rho}(q_0^2)},\\
&& D^{33}(q_0^2)=\frac{-2G_v}{1-2G_v\Pi^{33}_{\rho}(q_0^2)}.
\end{eqnarray}

In this way, the equations for the neutral $\rho$ mass associated to the three spin  projections ( $s_z=-1,1,0$) are given by
\begin{eqnarray}
&& 1-2G_v\Pi^{11}_{\rho}(q_0^2)=0, (s_z=\pm 1)  \, ,\\
&& 1-2G_v\Pi^{33}_{\rho}(q_0^2)=0, (s_z=0)  \,.
\end{eqnarray}
In the last equation we considered $\Pi^{11}_{\rho}(q_0^2)= \Pi^{22}_{\rho}(q_0^2)$.
\subsection{The regularization at eB=0} \label{reg_B0}
\subsubsection{Pauli-Villars regularization}
The polarization tensor given by Eq. (\ref{pol1}) at $eB=0$ will be represented by $\Pi^{\mu\nu}_{\rho}(q^2,eB=0)\equiv \Pi^{\mu\nu}_{\rho}(q^2,0)$ 
and it can be evaluated starting from the quark propagator in momentum space
\begin{equation}
 \tilde{S}(k)=\tilde{S}_{f}(k)=\frac{1}{{\slashed{k}-M + i\epsilon}}=
 \frac{\slashed{k}+M}{k^2-M^2 + i\epsilon} \; ,\label{prop0}
\end{equation}
by using standard quantum field theory methods. Hence, after the evaluation of the traces and considering the Feynman procedure for dealing with the 
product of propagators \cite{Itzy} the polarization tensor is given by
\begin{align}
\frac{1}{i} \Pi^{\mu\nu}_{\rho}(q^2,0)=&4N_cN_f \int_0^1dx\int\frac{d^4k}{(2\pi)^4}\frac{-2q^2x(1-x)}{(k^2-\overline{M}^2+i\epsilon)^2}\left(-g^{\mu\nu}+\frac{q^\mu q^\nu}{q^2}\right) \; ,
\end{align}
where $\overline{M}^2=M^2-x(1-x)q^2$. The transversality of the polarization tensor due to the Ward identity  is manifest in the previous expression. 
It is simple to see from power counting that this expression is logarithmically divergent and in this work we will adopt the Pauli-Villars
regularization scheme \cite{Davidson:1995fq}. The resulting regularized polarization tensor is
\begin{align}
\frac{1}{i}\Pi^{\mu\nu}_{\rho}(q^2,0)=&4N_cN_f\sum_{i}c_i\int_0^1dx\int\frac{d^4k}{(2\pi)^4}\frac{-2q^2x(1-x)}{(k^2-\overline{M}^2_i+i\epsilon)^2}\left(-g^{\mu\nu}+\frac{q^\mu q^\nu}{q^2}\right).  \label{pol_PV}
\end{align}
where $\overline{M}^2_i=M^2-x(1-x)q_0^2 +a_i\Lambda^2$, $c_i$ and $a_i$ are the Pauli-Villars coefficients and $\Lambda$ is the cutoff as discussed 
in Section \ref{formal0}. 

Since we are interested in the rho vector meson mass calculation, we restrict Eq. (\ref{pol_PV}) to the rho rest frame,  $\overrightarrow{q}=\overrightarrow{0}$ and 
$q_0=m_{\rho}$  and in this case one can easily show that the only non-null components of the polarization tensor are 
$\Pi^{11}_{\rho}=\Pi^{22}_{\rho}=\Pi^{33}_{\rho}$. After integrating in $dk^0$ the expression for the polarization function can be rewritten in the 
following simplified way
\begin{eqnarray}
\Pi^{11}_{\rho}(q_0^2,0)=\frac{N_c N_f q_0^2}{\pi^2}\sum_{i}c_i\int_0^1dx\int_{0}^{\infty}\frac{dk \; k^2 x(1-x)}{(k^2+\overline{M_i}^2-i\epsilon)^{3/2}} \; .
\label{piB0}
\end{eqnarray}
In the latter expression the integral in $dk$ can be calculated analytically yielding the regularized expression for the polarization $\Pi^{11}_{\rho}$
\begin{eqnarray}
\Pi^{11}_{\rho}(q_0^2,0)=-\frac{N_cN_fq_0^2}{2\pi^2}\int_0^1dx\;  x(1-x)\left[\log\left(1+\frac{2\Lambda^2}{\overline{M}^2}\right)-2\log\left(1+\frac{\Lambda^2}{\overline{M}^2}\right)\right]. 
\label{POl_B0_PV}
\end{eqnarray}
This expression will be used in our numerical calculations as discussed in Section \ref{results}.

\subsection{The tensor $\Pi^{\mu\nu}_\rho$ with spin  projection $s_z=1$ at $eB\neq 0$}
In this section we will discuss the main steps for the evaluation of the component $\mu=\nu=1$ of the polarization tensor $\Pi^{\mu\nu}$ in the case of 
an external magnetic field. After substituting in Eq. (\ref{pol1}) the fermion propagator in the Schwinger form, Eq. (\ref{schwinger}), one obtains
\begin{align}
 &\Pi^{11}_{\rho}(q^2,eB)= - i N_c \sum_{f=u,d}\int\frac{d^4k}{(2\pi)^4}  
 \int_0^{\infty} ds \int_0^{\infty}dt  e^{is(k_{\parallel}^2-k_{\perp}^2\frac{\tan \beta_fs}{\beta_fs}-M^2)}
 \; e^{it(k_{\parallel}^{\prime 2}-k_{\perp}^{\prime 2}\frac{\tan \beta_ft}{\beta_ft}-M^2)}
  \tr_{D}[\gamma^{1}\hat{D_f}(k,s)\gamma^{1}\hat{D_f}(k^\prime,t)]. 
 \label{pi_11}
\end{align}
In the last expression, $k^\prime \equiv k+q$ and $\hat{D_f}(k,s)$  is the matricial part of $\tilde{S}_f(k)$. In the following, we present the results obtained 
for the latter trace in the rest frame of the rho meson
\begin{align}
\tr[\gamma^{1}\hat{D_f}(k,s)\gamma^{1}\hat{D_f}(k^\prime,t)] = 
4 \left[(k_0k_0'-k_3^2-M^2)h_f(s,t)+(k_1^2-k_2^2)g_f(s)g_f(t)\right] \; ,
\label{trace_11}
\end{align}
where we have defined $h_f(s,t)=1+\tan(\beta_fs)\tan(\beta_ft)$ and $k_0'=k_0+q_0$. The evaluation of the previous integral is simpler in 
Euclidean coordinates, so we apply the transformation to the Euclidean plane ($k_0\rightarrow ik_4$, $q_0\rightarrow iq_4$). 
After substituting the trace given in Eq. (\ref{trace_11}) and performing the change of coordinates ($s\rightarrow -is$, $t\rightarrow -it$) 
in Eq. (\ref{pi_11}) one obtains \cite{Das:2019ehv}
%Using the definition,
%$\Pi^{11,33}_{\rho}=\sum_{f=u,d}\Pi^{11}_{\rho}$, we can write
\begin{align}
 \Pi^{11}_{\rho}(q_4^2,eB)=\sum_{f=u,d}\Pi^{11}_{\rho}(q_4^2,\beta_f) \; ,
\end{align}
where 
\begin{align}
 &\Pi^{11}_{\rho}(q_4^2,\beta_f)=\int_k\int_s\int_t  e^{-[ sk_4^2+t(k_4+q_4)^2+\omega k_3^2+\Omega_f k_{\perp}^2+\omega M^2]} \nonumber\\
 &\times(-4)N_c\left[-\left(k_4 (k_4+q_4)+k_3^2+M^2\right)\tilde{h}_f(s,t)  +(k_1^2-k_2^2)\tilde{g}_f(s)\tilde{g}_f(t)\right], \label{pi_11_int}  
\end{align}
and the following shorthand notations have been adopted: 
%$\int_k=\frac{1}{(2\pi)^4}\int dk_1 dk_2 dk_3 dk_4$, $\int_t\int_s=\int_0^{\infty}dt%\int_0^{\infty}ds$,
%$\omega=(s+t)$, $\Omega_f(s,t)=\frac{\tanh(\beta_ft)+\tanh(\beta_fs)}{\beta_f}$, 
%\tilde{h}_f(s,t)=1-\tanh(\beta_fs)\tanh(\beta_ft)$ and $\tilde{g}_f(s)=1-\tanh ^2 (\beta_f s)$.
%
\begin{align}
&\int_k \equiv \frac{1}{(2\pi)^4}\int dk_1 dk_2 dk_3 dk_4 \; ,  \; \int_t\int_s\equiv \int_0^{\infty}dt\int_0^{\infty}ds \; ,  \nonumber \\
&\omega=(s+t) \; , \; \Omega_f(s,t)=\frac{\tanh(\beta_fs)+\tanh(\beta_ft)}{\beta_f} \; , \nonumber \\ 
&\tilde{h}_f(s,t)=1-\tanh(\beta_fs)\tanh(\beta_ft) \; , \nonumber \\
& \; \tilde{g}_f(s)=1-\tanh ^2 (\beta_f s) \; .
\end{align}
The next step involves to perform  a  series of Gaussian integrals in the momentum space in Eq. (\ref{pi_11_int}), which results in
\begin{align}
 \Pi^{11}_{\rho}(q_4^2,\beta_f)=\frac{N_c}{4\pi^2}\int_s \int_t  \frac{e^{-\omega {\cal M}^2_+}}{\omega}\left(\frac{1}{\omega}+{\cal M}^2_-\right)
 \frac{\tilde{h}_f(s,t)}{\Omega_f(s,t)},
\end{align}
where $ { \cal M}^2_\pm \equiv M^2 \pm \frac{q_4^2}{4}(1-\chi^2)$ and $\chi \equiv \frac{t-s}{t+s}$. After performing the change of variables  
$\beta_fs\rightarrow s'$ and $\beta_ft\rightarrow t'$ in the latter integral, we obtain
\begin{align}
 \Pi^{11}_{\rho}(q_4^2,\beta_f)=&\frac{N_c\beta_f}{4\pi^2}\int_0^{\infty} ds 
 \int_0^{\infty} dt \;
 \frac{e^{-\omega \frac{{\cal M}^2_+}{\beta_f}}}{\omega} 
  \left(\frac{1}{\omega}+\frac{{\cal M}^2_-}{\beta_f}\right)
                           \left(\frac{1-\tanh s \tanh t}{\tanh s+\tanh t}\right),
                           \label{int_pi_11_5}
\end{align}
in the latter integral the original dummy variables of integration $s$ and $t$ have been recovered.
Now, we introduce for convenience a new set of variables $v$ and $x$ through the relations: $s=v(1-x)$ and $t=vx$. 
The integral over $s$ and $t$ from  $[0,\infty)$in Eq. (\ref{int_pi_11_5}) now becomes an integral over $v$ and $x$ where
$v \in [0,\infty)$ and $x \in [0,1]$. Noting that the Jacobian of the transformation is $v$, we obtain
\begin{align}
 \Pi^{11}_{\rho}(q_0^2,\beta_f)=&\frac{N_c\beta_f}{4\pi^2}\int_0^{1} dx \int_0^{\infty} dv \;
 e^{-v \frac{\bar{\cal M}^2_+}{\beta_f}} 
 \left(\frac{1}{v}+\frac{\bar{\cal M}^2_-}{\beta_f}\right) \Phi(v,x)
                         \;  ,  \label{pi_11_unreg}
\end{align}
where
\begin{align}
 &\Phi(v,x) =\left\{\frac{1-\tanh\left[v(1-x)\right] \; \tanh \left(vx\right)}{\tanh \left[v(1-x)\right] +\tanh \left(vx\right)}\right\} \nonumber \; ,\\
 &\bar{\cal M}_{\pm}^2=M^2 \pm q^2_4 \; x(1-x) = M^2 \mp  \; x (1-x) \; q_0^2 .
 \label{M_pm}
\end{align}
For the calculation of the rho meson mass, we have reintroduced the Minkowski component  $q_0^2=-q_4^2$ since $q_0$=$m_\rho$.

% -----------------------------------------------------------------
%
\subsubsection{Regularization of $\Pi^{11}_\rho$}
%
% ------------------------------------------------------------------
%
%
As already discussed in Section \ref{Introdu} the regularization procedure is a crucial issue for obtaining reliable results from the effective model 
calculations under strong external magnetic fields. In this work, we use the MFIR prescription to regularize the Eq. (\ref{pi_11_unreg}) \cite{Avancini:2019wed}.
The MFIR procedure corresponds to the separation of the explicit finite magnetic contribution from the infinite non-explicit magnetic-dependent contribution. 
This latter contribution is to be identified with the usual one that appears in the NJL model without an external magnetic field. It is worth to mention that a similar
procedure was used long ago in the context of the electromagnetic response in strong magnetic fields \cite{Bakshi:1976}. We first rearrange 
$\Pi^{11}_\rho(q_0^2,\beta_f)$ by subtracting $\Pi^{11}_{\rho}(q_0^2,\beta_f=0)$ yielding the finite magnetic contribution: 
\begin{align}
  &\Pi^{11}_{\rho,F}(q_0^2,\beta_f) \equiv \Pi^{11}_{\rho}(q_0^2,\beta_f) - \Pi^{11}_{\rho}(q_0^2,\beta_f=0)   \; . \label{finite_Pi_11}
\end{align}
Next, we add to this expression the already regularized $eB=0$ term given in Eq. (\ref{POl_B0_PV}) of Section-\ref{reg_B0} and thus we obtain the regularized 
$\Pi^{11}_{\rho}$
\begin{align}
 \Pi^{11}_{\rho,R}(q_0^2,\beta_f) =  \Pi^{11}_{\rho,F}(q_0^2,\beta_f=0)  
 + \Pi^{11}_{\rho}(q_0^2,0) \; .
\end{align}

The source of the divergence in Eq. (\ref{pi_11_unreg}) occurs when $v$ goes to zero. 
To obtain the finite magnetic contribution, Eq. (\ref{finite_Pi_11}), we start by calculating the limit $\beta_f \to 0$ in Eq. (\ref{pi_11_unreg}).  
This limit is easier to calculate making the change of variable $v \to \beta_f y$ in Eq. (\ref{pi_11_unreg}), where $\Pi^{11}_{\rho}(q_0^2,\beta_f)$ 
assumes the form
\begin{align}
 &\Pi^{11}_{\rho}(q_0^2,\beta_f)=\frac{N_c}{4\pi^2}\int_0^{1} dx \int_0^{\infty} dy \;
 e^{-y \bar{\cal M}^2_+}  
 \left(\frac{1}{y}+\bar{\cal M}^2_-\right) \beta_f \Phi(\beta_f y,x)
                         \;  ,  \nonumber
\end{align}
expanding the function $\Phi(\beta_f y,x)$ in a Taylor series expansion about 
$\beta_f y\sim 0$: 
\begin{eqnarray}
 &&\Phi(\beta_f y,x)=\left\{\frac{1-\tanh \left[\beta_f y (1-x)\right] \; \tanh \left(\beta_f y x\right)}
 {\tanh \left[\beta_f y (1-x)\right] + \tanh \left(\beta_f y x\right)}\right\}  ,\nonumber\\
 &&\approx \frac{1}{\beta_f y}+ \left( \frac{1}{3} - 2x(1-x) \right) \beta_f y  + 
 \mathcal{O}((\beta_f y)^3) \; , \; (v \ll 1) \; ,\label{expand} \nonumber\\
\end{eqnarray}
\ni 
we now obtain as the result of the limit
\begin{align}
  \Pi^{11}_{\rho}(q_0^2,\beta_f=0) &= \lim_{\beta_f \to 0} \Pi^{11}_{\rho}(q_0^2,\beta_f) 
 \nonumber \\
 &= \frac{N_c}{4\pi^2}\int_0^{1} dx \int_0^{\infty} dy \;
 e^{-y \bar{\cal M}^2_+} 
   \left(\frac{1}{y}+\bar{\cal M}^2_-\right) \frac{1}{y} \; . \label{Pi^11_B0}
\end{align}
As expected this expression is infinity and does not depend explicitly on $eB$. It is possible to rearrange this latter expression and to show that it can be written in the form given in  Eq. (\ref{piB0}). Putting this expression in Eq. (\ref{finite_Pi_11}) we obtain the finite pure magnetic term of $\Pi^{11}_{\rho}$
\begin{align}
 \Pi^{11}_{\rho,F}(q_0^2,\beta_f)=&\frac{N_c\beta_f}{4\pi^2}\int_0^{1} dx \int_0^{\infty} dv \;
 e^{-v \frac{\bar{\cal M}^2_+}{\beta_f}} 
 \left(\frac{1}{v}+\frac{\bar{\cal M}^2_-}{\beta_f}\right) \Phi(v,x)_R
                         \;  ,  \label{pi_11_reg}
\end{align}
where we define 
\begin{eqnarray}
 \Phi_R(v,x)=\left\{\frac{1-\tanh \left[v(1-x)\right] \tanh \left(vx\right)}{\tanh \left[v(1-x)\right]+\tanh \left(vx\right)}-\frac{1}{v}\right\},
\end{eqnarray}
It is evident that the MFIR procedure remove the divergence subtracting the singularity. The final regularized expression for $\Pi^{11}_{\rho}$ is now given by
\begin{eqnarray}
 \Pi^{11}_{\rho,R}(q_0^2,eB)=\Pi^{11}_{\rho}(q_0^2,0) + \sum_{f=u,d}\Pi^{11}_{\rho,F}(q_0^2,\beta_f),\label{finalsz1}
\end{eqnarray}

\noindent where $\Pi^{11}_{\rho}(q_0^2,0)$ is the vacuum contribution given in 
Eq. (\ref{POl_B0_PV})  that depends on the regularization prescription adopted. 

\subsection{Tensor $\Pi^{\mu\nu}_\rho$ with spin  projection $s_z=0$ at $eB\neq 0$}
%
% -----------------------------------------------------------------------------------------
%
%
We discuss briefly here how to evaluate the $s_z=0$ polarization function by using exactly the same techniques of the last section. We start by substituting the fermion propagator, Eq. (\ref{schwinger}), in the component 
$\mu=\nu=3$  of the polarization tensor $\Pi^{\mu\nu}$ given in Eq. (\ref{pol1}), 
obtaining
\begin{align}
 \Pi^{33}_{\rho}(q_0^2,eB)= - i N_c \sum_{f=u,d}\int\frac{d^4k}{(2\pi)^4}  
 \int_0^{\infty} ds \int_0^{\infty}dt  e^{is(k_{\parallel}^2-k_{\perp}^2\frac{\tan \beta_fs}{\beta_fs}-M^2)}
 \; e^{it(k_{\parallel}^{\prime 2}-k_{\perp}^{\prime 2}\frac{\tan \beta_ft}{\beta_ft}-M^2)}
  \tr_{D}[\gamma^{3}\hat{D_f}(k,s)\gamma^{3}\hat{D_f}(k^\prime,t)]. 
 \label{pi_33}
\end{align}
\noindent where $k^\prime = k+q$ and again we restrict the calculation to the rest frame of the rho meson. The explicit calculation of the trace results in

\begin{align}
\tr_{D}[\gamma^{3}\hat{D_f}(k,s)\gamma^{3}\hat{D_f}(k+q,t)]=4\left[(k_0k_0'+k_3^2-M^2)n_f(s,t)+(k_1^2+k_2^2)g_f(s)g_f(t)\right]  \; ,
\end{align}

\noindent where $n_f(s,t)\equiv 1-\tan(\beta_fs)\tan(\beta_ft)$. 

\ni Following the same steps of the last section, all the Gaussian integrals can be performed and making the same change of variables as before one obtains:
\begin{align}
 \Pi^{33}_{\rho}(q_0^2)&=\sum_{f=u,d}\Pi^{33}_{\rho}(q_0^2,\beta_f) \nonumber \; , \\
 \Pi^{33}_{\rho}(q_0^2,\beta_f)&=\frac{N_c\beta_f^2}{4\pi^2}\int_{0}^1dx \int_0^{\infty}dy \; e^{-y \bar{\cal M}_+^2 } \left[\frac{\bar{\cal M}_-^2}{\beta_f}\coth (\beta_f y) +\sinh^{-2} (\beta_f y) \right] \; ,
\end{align}
where $\bar{\cal M}_\pm$ was defined in Eq. (\ref{M_pm}).
For the regularization of the latter expression, we use the same reasoning of the previous section. First, we consider the following Taylor series expansion of $\coth v$ and 
$1/\sinh^2 v$ about $v=0$: 
\begin{align}
&\coth v \sim \frac{1}{v}+\frac{v}{3}+\mathcal{O}(v^3) \nonumber \; ,\\ 
&\sinh^{-2} v\sim \frac{1}{v^2}-\frac{1}{3}+\mathcal{O}(v^2)  
\end{align}
From these series expansions is easy to calculate the limit
\begin{align}
 & \Pi^{33}_{\rho}(q_0^2,\beta_f=0) = \lim_{\beta_f \to 0} \Pi^{33}_{\rho}(q_0^2,\beta_f) 
 \nonumber \\
 &= \frac{N_c}{4\pi^2}\int_0^{1} dx \int_0^{\infty} dy \;
 e^{-y \bar{\cal M}^2_+} 
   \left(\frac{1}{y}+\bar{\cal M}^2_-\right) \frac{1}{y} \; . \label{Pi^33_B0}
\end{align}
We notice that this latter expression coincides with Eq. (\ref{Pi^11_B0}) obtained in previous section for the 
$\Pi^{11}_{\rho}(q_0^2,\beta_f=0)$ component of the polarization tensor. Of course, that should be expected for symmetry reasons, since for $B=0$ there is no preferred direction in space. 
\noindent Therefore, as in Eq. (\ref{finalsz1}) the final regularized expression for $\Pi^{33}_{\rho}$ is now given by
\begin{eqnarray}
 \Pi^{33}_{\rho,R}(q_0^2,eB)=\Pi^{33}_{\rho}(q_0^2,0) + \sum_{f=u,d}\Pi^{33}_{\rho,F}(q_0^2,\beta_f)  \label{finalsz0} \; ,
\end{eqnarray}
 where again $\Pi^{33}_{\rho}(q_0^2,0)$ is the vacuum contribution given in 
Eq. (\ref{POl_B0_PV}) and the finite magnetic contribution is 
\begin{eqnarray}
\Pi^{33}_{\rho,F}(q_0^2,\beta_f)=\frac{N_c\beta_f}{4\pi^2}\int_{0}^1dx \int_0^{\infty}dv \; e^{-v \frac{\bar{\cal M}^2_+}{\beta_f}}\left[\frac{\bar{\cal M}^2_-}{\beta_f}\left(\coth v-\frac{1}{v}\right)+\left(\frac{1}{\sinh^{2} v}-\frac{1}{v^2}\right) \right].
\end{eqnarray}

Once the formalism has been presented, we are now in condition to perform 
a numerical analysis. The NJL model and its extensions presents itself as one of the simplest and successful methods to evaluate light meson properties in strong magnetic fields (e.g. the masses, decay width, effective couplings etc). There are a great number of examples recently published such as the evaluation of the neutral $\pi^0$ and $\sigma$ meson properties in Refs. \cite{Avancini:2016fgq,Avancini:2015ady,Avancini:2018svs}, charged $\pi$ meson \cite{Coppola:2019uyr,Coppola:2018vkw} with the $N_f=2$ version of the model and more recently in the $N_f=3$, in which several meson masses with strange quark content are evaluated \cite{Avancini:2021pmi}.
Most of the mismatch between the NJL model and thermo-magnetic effects 
are due to the phase transition that occurs when the temperature increases and exceeds a
pseudo-critical value. Since in this work we are constrained to zero temperature, we decided
to perform the calculation of the $\rho$-meson mass within the conventional constant-coupling
NJL model.

%
%
% -------------------------------------------------------------
%
\section{Numerical Results}\label{results}
%
% -------------------------------------------------------------
%
%
In this section we will present our numerical results. First, the set of parameters of the Pauli-Villars parametrization are chosen as $\Lambda=761.22$ MeV, 
$m_0=6.565$ MeV and $G=3.5758/\Lambda^2$. This choice of parameters complies with the vacuum values of the condensate, 
$\left \langle \overline{u}u \right \rangle^{1/3}=-250~{\rm MeV}$, the  pion decay constant, $f_{\pi}=107$ MeV 
and the pion mass, $m_{\pi}=135$ MeV. The $\rho$-meson mass $m_{\rho}=744.772$ MeV is obtained at $eB=0$ for $G_v=1.3\;G$.

\begin{figure}[H]
    \centering
    \includegraphics[width=0.6\textwidth]{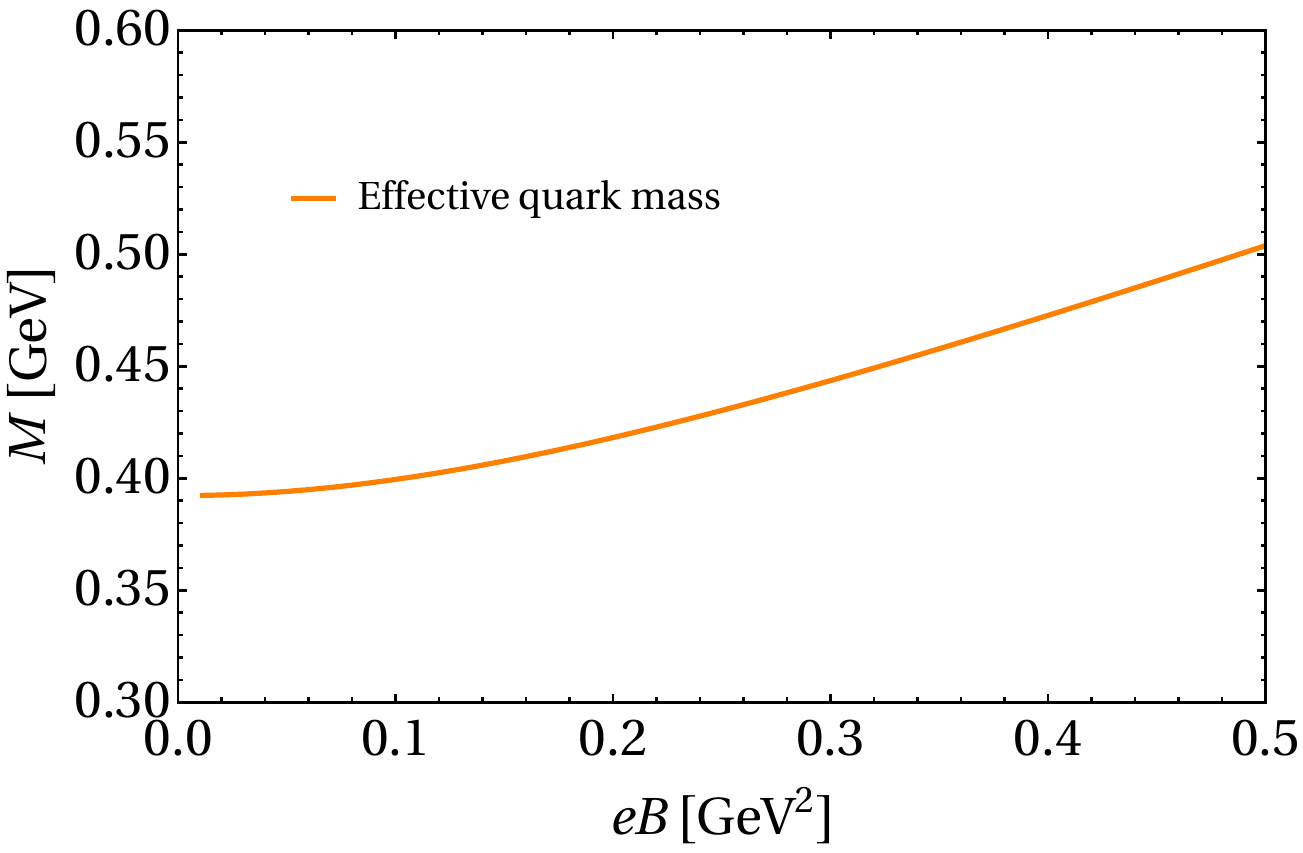}
    \caption{The effective quark mass  as a function of the magnetic field.}
    \label{effmass}
\end{figure}

In Figure \ref{effmass} we show the well known effective quark mass as a function of the magnetic field, showing the expected magnetic catalysis effect. 

\begin{figure}[H]
    \centering
    \includegraphics[width=0.6\textwidth]{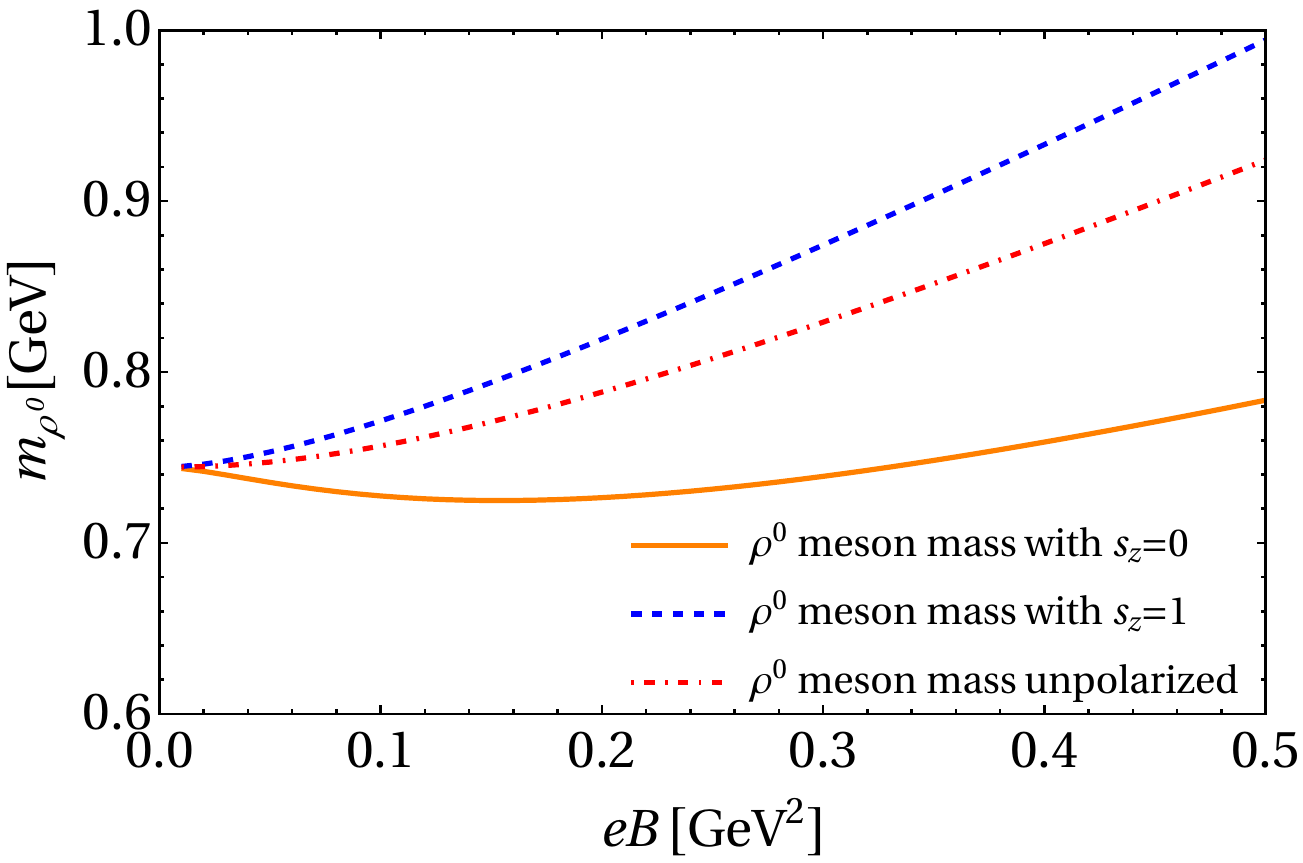}
    \caption{Neutral $\rho$ meson mass with different spin projections at finite magnetic field in the two-flavor ($N_f=2$) NJL model.}
    \label{meson}
\end{figure}

With the obtained values of effective quark mass we can calculate the mass of different projections of the $\rho^0$ meson, $s_z=0,\pm 1$, as we show 
in Fig. \ref{meson}. In the same figure we obtain the unpolarized $\rho^0$ meson mass in the same way as defined in \cite{Luschevskaya:2014lga}, in which 
we have $m_{\rho^0}=(m_{\rho^0_{s_z=1}}+m_{\rho^0_{s_z=-1}}+m_{\rho^0_{s_z=0}})/3$. We can see that the mass with spin projection $s_z=\pm 1$ is
catalyzed as the magnetic field increases. Nevertheless, the mass with $s_z=0$ decreases until $eB\sim 0.15$ GeV$^2$ and passes to increases weakly. As expected, 
the unpolarized mass is catalyzed as the magnetic field increases as an effect of the $s_z=\pm1$ spin projection that has weight $2/3$ of the unpolarized 
$\rho^0$ mass calculation.

\begin{figure}[H]
    \centering
    \includegraphics[width=0.6\textwidth]{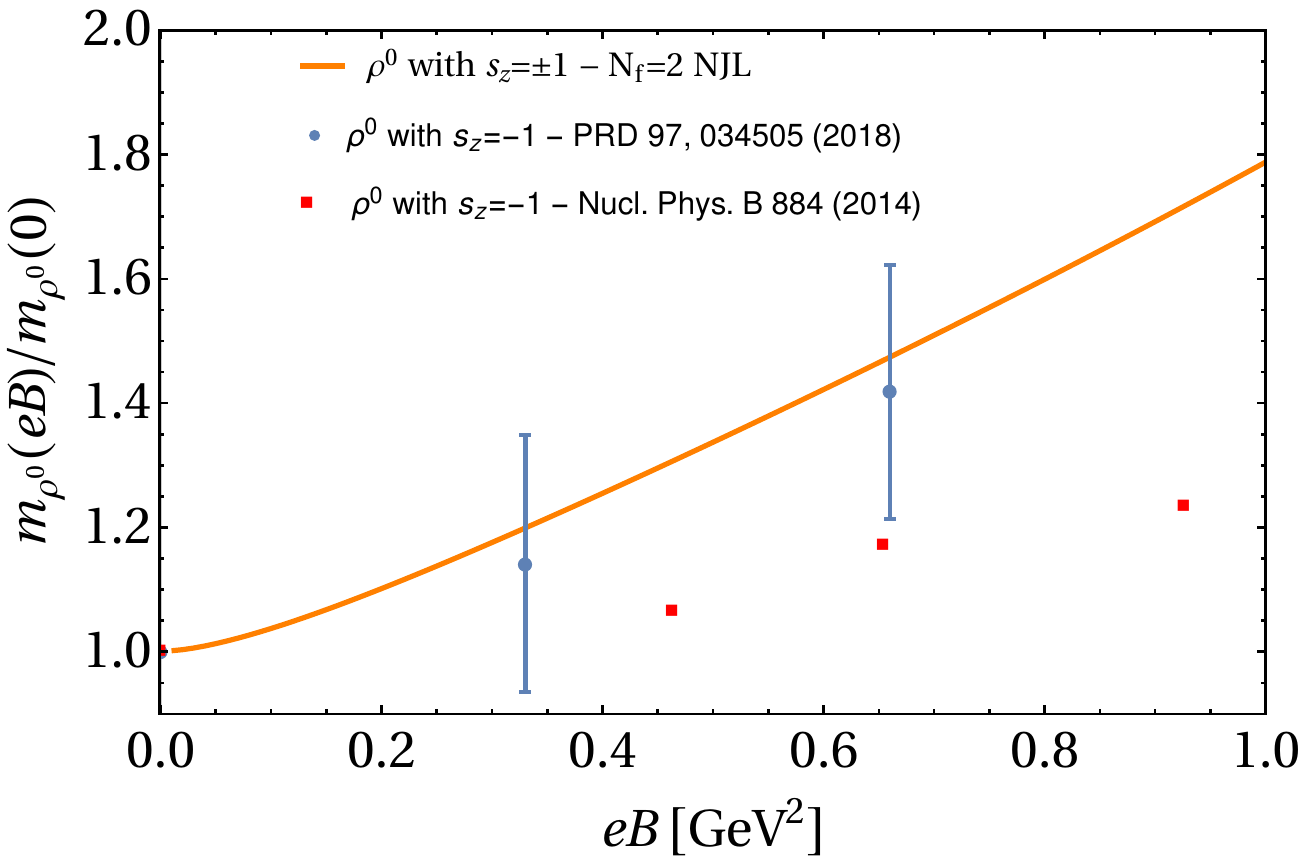}
    \caption{Neutral $\rho$ meson mass with spin projection $s_z=\pm1$ at finite magnetic field in the two-flavor ($N_f=2$) NJL model compared with LQCD  results\cite{Bali:2017ian,Luschevskaya:2012xd}.}
    \label{mesonsz1LQCD}
\end{figure}

We can compare some of our $s_z=\pm1$ spin projection mass with different results from LQCD in the quenched approximation. In Fig. \ref{mesonsz1LQCD}, 
we can observe that the results of normalized $\rho^0$ meson mass by its vacuum value mass are in good agreement in what is expected, quantitatively 
with Ref. \cite{Bali:2017ian} and qualitatively with Ref.     \cite{Luschevskaya:2012xd}. In Refs. \cite{Andreichikov:2016ayj,Andreichikov:2013zba} we also observe the projection $s_z=\pm 1$ of the $\rho^0$ mass increasing as a function of the magnetic field.

The mass of the $s_z=0$ projection, on the other hand, is in a completely different trend when compared to the available lattice QCD results
\cite{Luschevskaya:2012xd}, in which the mass decreases as function of the magnetic field. As explained in ref. \cite{Bali:2017ian}, the $s_z=0$ 
projection can present effects of vacuum diagrams, which are not easy to handle in LQCD calculations. Furthermore, the lattice data used in this work are from two different groups which uses different approaches. In ref. \cite{Bali:2017ian} quenched Wilson fermions are used, while overlap fermions are applied in Ref. \cite{Luschevskaya:2012xd} and, in both methods, problems with mixing states of $\pi$ and $\rho$ meson may arise. For these reasons it is difficult, at the moment, to obtain a definitive answer for the preliminary contrast between lattice and model results in the $s_z=0$ projection. Therefore, these comparisons can be reconsidered 
in future LQCD simulations. Also, the results obtained by Refs. \cite{Andreichikov:2016ayj,Andreichikov:2013zba} show the mass of the $s_z=0$ projection of the $\rho^0$ meson increasing as a function of the magnetic field.

\begin{figure}[H]
    \centering
    \includegraphics[width=0.6\textwidth]{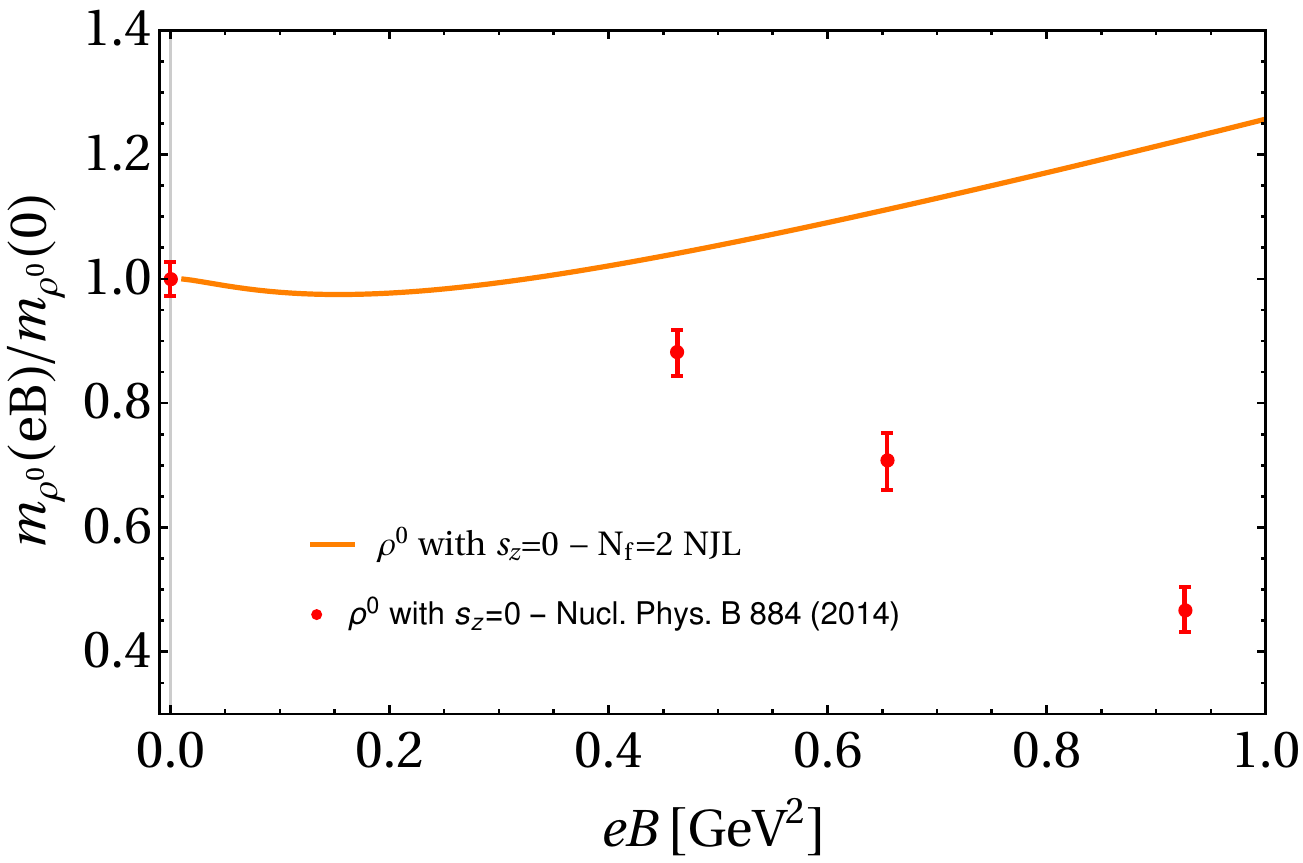}
    \caption{Neutral $\rho$ meson mass with spin projection $s_z=0$ at finite magnetic field in the two-flavor ($N_f=2$) NJL model compared with LQCD results \cite{Luschevskaya:2012xd}.}
    \label{fig:rmsz1}
\end{figure}

\begin{figure}[H]
    \centering
    \includegraphics[width=0.6\textwidth]{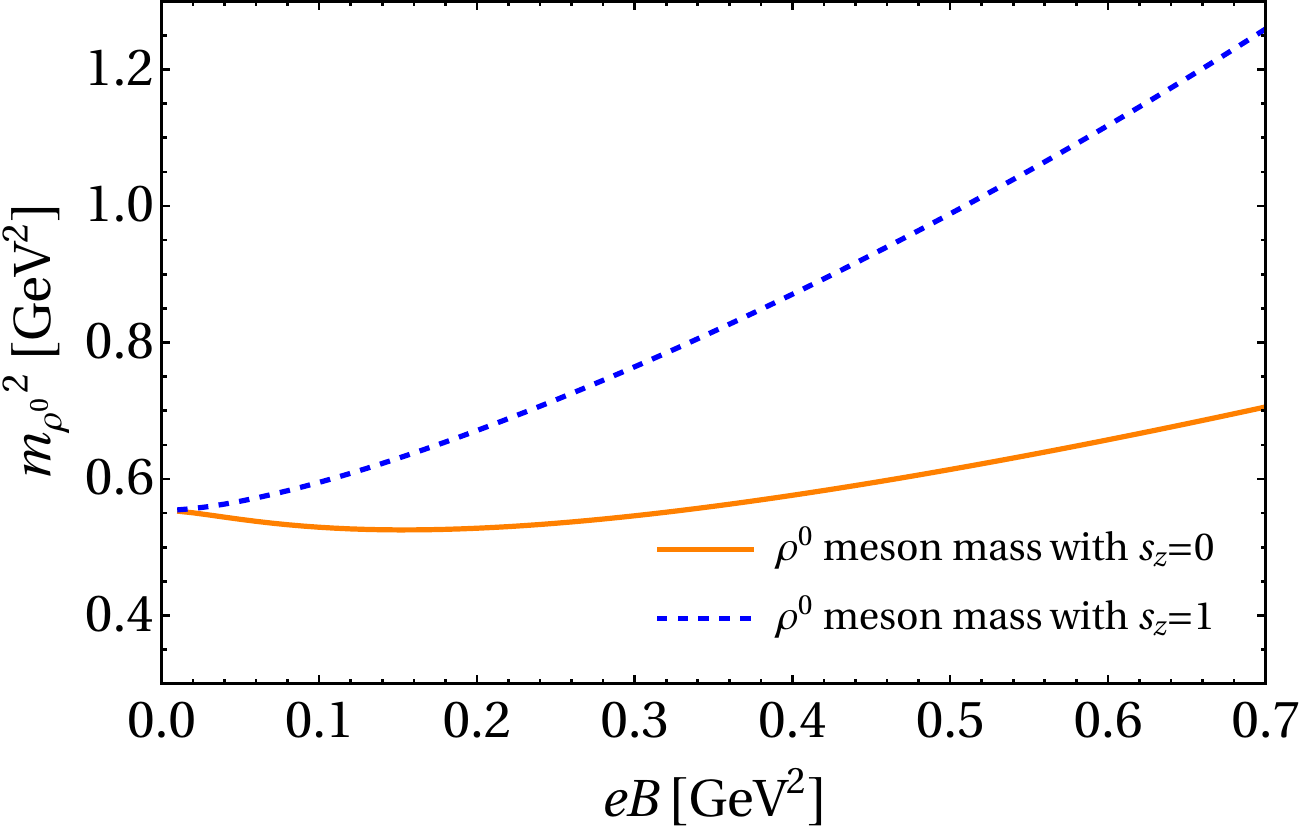}
    \caption{Squared neutral $\rho$ meson mass with spin projections $s_z=0$ and $s_z=1$ at finite magnetic field in the two-flavor ($N_f=2$) NJL.}
    \label{rhosquared}
\end{figure}

Moreover, we also show in Fig. \ref{rhosquared} the squared values of the $\rho^0$ meson mass. We see that the behavior of both projections follows 
what we have seen in Fig. \ref{meson}, but they don't present the rapid growth as observed in Ref. \cite{Liu:2014uwa}. Besides, we also don't 
observe any oscillation in the masses with the three spin projections, $s_z=0,\pm 1$, we have explored. This is mainly due the regularization prescription based in MFIR procedure, 
as explored in detail in Ref. \cite{Avancini:2019wed}.

\begin{figure}[H]
    \centering
    \includegraphics[width=0.6\textwidth]{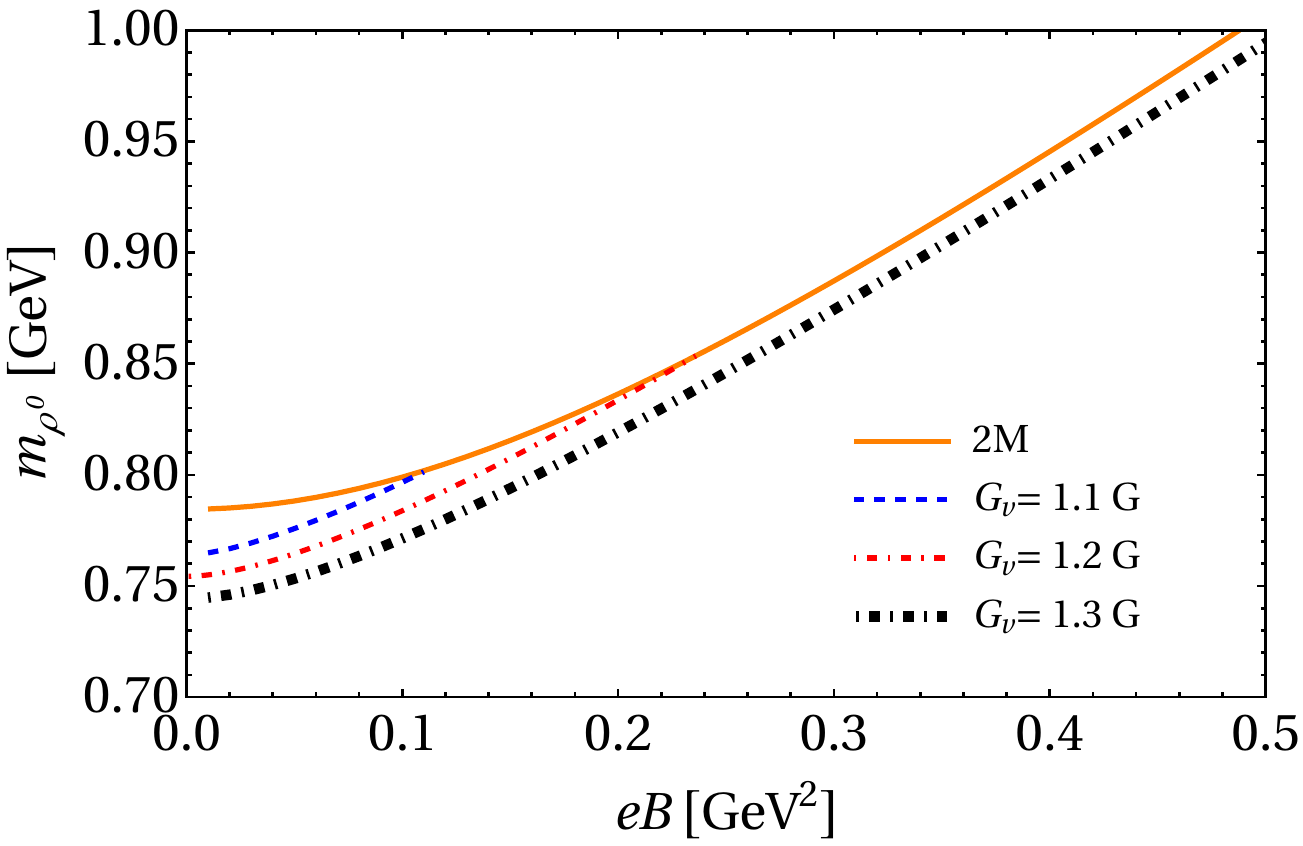}
    \caption{Neutral $\rho$ meson mass with spin projection $s_z=\pm1$ at finite magnetic field in the two-flavor ($N_f=2$) NJL model with different vector couplings 
    compared to $2M$.}
    \label{rhoGVs}
\end{figure}

In order to compare our results for the $\rho^0$ mass with the ones that have been obtained in the literature 
 in both
effective model approaches and LQCD evaluations, we explore in our calculations the parametrizations where the $\rho^0$
meson mass solution is a stable particle rather than a resonance \cite{Bali:2017ian}.
 A straightforward analysis of the polarization tensor shows that a stable particle solution is obtained whenever the threshold relation $m_{\rho^0}^2<4M^2$ is satisfied.
 As it is well known, the effective quark mas $M$ suffers magnetic catalysis, as can be seen in Fig. \ref{effmass}. Hence, since the $\rho^0$ mass decreases for the spin projection $s_z=0$ as a function of $eB$ (see Fig. \ref{meson}) the threshold condition will be respected in this case, at least, in the magnetic field range that we are considering in this work.
 In contrast to the $s_z=0$ spin projection, it is clear from the Fig. \ref{meson} that the $\rho^0$ meson mass with spin projections $s_z=\pm 1$  is
always catalyzed as a function of the magnetic field and due to this behavior, the threshold condition just mentioned it is possibly not satisfied and thus induce a resonant state solution depending on the value of the free parameter $G_v$.
 In order to  understand when the resonant solutions appear, we study how the  $\rho^0$ meson mass with $s_z=\pm1$ spin projection depends on the vector coupling strength $G_v$. Our results are shown in Fig. \ref{rhoGVs}, where $\rho^0$ meson mass are 
calculated using three different sets of vector couplings, $G_v=1.1\;G$, $G_v=1.2\;G$ and $G_v=1.3\;G$. Note that the values of $2M$ are shown to facilitate the discussion. It is clear from the latter figure that the threshold condition $m_{\rho^0}^2<4M^2$ will be satisfied until $eB\sim 0.11$ GeV$^2$ in the case of $G_v=1.1\;G$, $eB\sim 0.24$ GeV$^2$ for $G_v=1.2\;G$ and for $G_v=1.3\;G$  bound state solutions are found in the whole range of $eB$ considered.
However, the choice of $G_v$ has to be taken with care in order not to lose
consistency, once we can clearly see that the effective quark mass doesn't reproduce the vacuum value for the $\rho$ meson $m_{\rho^0}=775.49$ MeV.

\begin{figure}[H]
    \centering
    \includegraphics[width=0.6\textwidth]{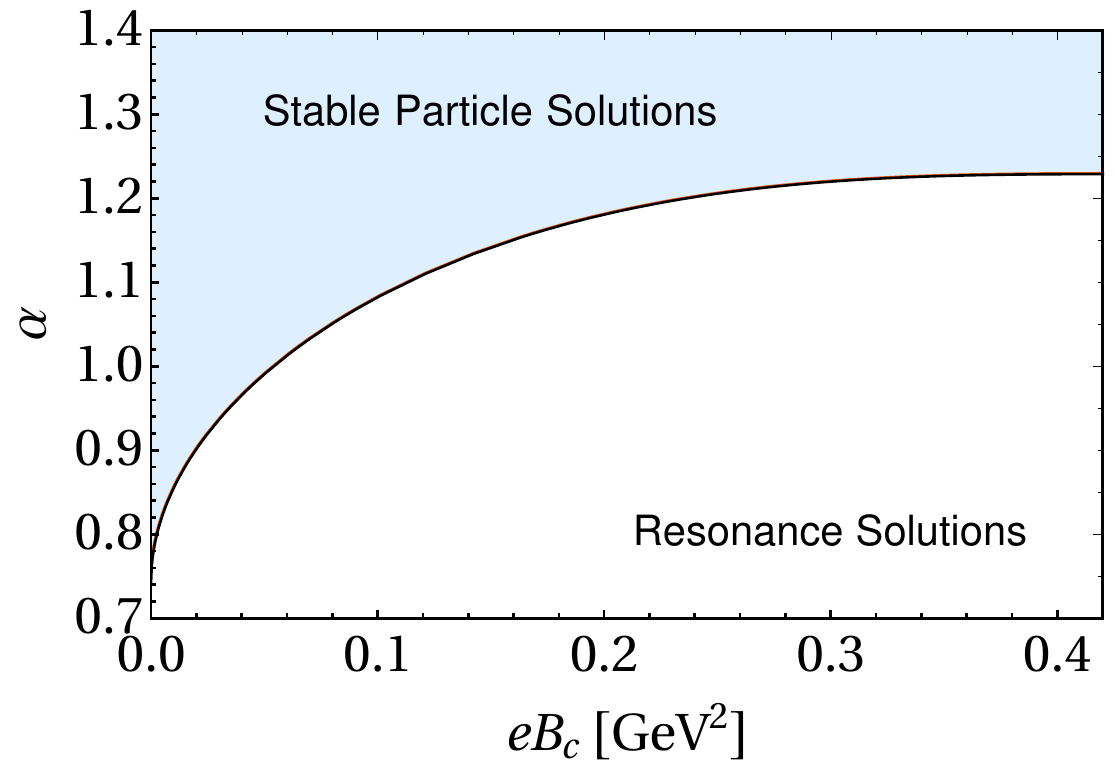}
    \caption{The proportionality factor between the vector and the scalar couplings, 
    $\alpha = G_v / G$, as a function of the critical magnetic field $eB_c$.}
    \label{Gvs}
\end{figure}

 Here, we define the critical magnetic field, $eB_c$, as the value of $eB$ at which the bound state turns out to be a resonant state.
 Aiming to determine precisely the $G_v$ parameter range assuring that the $\rho^0$ meson mass is not a resonant state, we define the vector coupling constant as a product of a proportionality factor $\alpha$ times the scalar coupling constant $G$, i. e., 
$G_v \equiv \alpha \times G$.  In Fig. \ref{Gvs}, we show the
proportionality factor $\alpha$ as function of the critical magnetic field, where the region below the line corresponds to resonant state solutions for the $\rho$ mass.
We conclude that to obtain in a wider range of magnetic fields bound states  
solutions for the $\rho^0$ meson mass with $s_z=\pm1$ , e.g. $eB\gtrsim0.4$ GeV$^2$ 
one should adopt $\alpha\gtrsim 1.3$.
Once the strong magnetic fields generated in peripheral heavy ion
collisions can induce magnetic fields of the order $eB\sim 0.2$ GeV$^2$, and the range of applicability of the NJL model is $eB\sim\Lambda^2\sim 0.580$
GeV$^2$, we can safely work with a more convenient value, i.e., $\alpha<1.3$.

\section{Conclusions}\label{conclusions}
So far, we have fully evaluated the polarization tensor functions of the vector channel and obtained the neutral $\rho^0$ meson with three different spin projections, $s_z=0,\pm 1$ in the context of the magnetized two flavor NJL model. To this end, we have employed the MFIR procedure, in order to avoid 
non-physical oscillations previously observed in the literature with non-MFIR methods.

 Instead of using propagators with Landau level summation, we use propagators in the Schwinger proper-time formalism, where the summation over the Landau levels is analytically circumvented and clear advantages follow from that. Starting from the non-regularized expressions, the limit $eB \to 0$ which is difficult to be performed using Landau summation is, nevertheless, straightforward in the PT formalism.  To separate the  non-explicitly $eB$-dependent divergent contribution from the finite $eB$ dependent term is straightforward as discussed in detail in the manuscript. 
So our expressions are valid for all the $eB$ range  and this has immediate consequences concerning the regularization.
To our knowledge, there is not any paper in the literature where these issues are treated with the details that we have used, and, certainly, this may be useful for other authors exploring the same kind of calculations.

 Our results show that the $\rho^0$ meson mass  with spin projection $s_z=\pm 1$ increases as a function of the magnetic field. On the other hand the $\rho^0$ meson mass with spin projection $s_z=0$ has a non-monotonic behavior. We find a decrease of the $\rho$ mass until $eB\sim 0.15$ GeV$^2$ and a weak increase after this value  with the increasing of the magnetic field. The spin projections of neutral $\rho$ mass are also compared with different sets of data from two lattice QCD groups. The $s_z=\pm 1$ case shows quantitatively and qualitatively good agreement with the two sets. In the $s_z=0$ case, we can see that the available LQCD data predicts a decrease of the neutral $\rho$ mass as a function of the magnetic field, which is in clear contrast with the present two-flavor ($N_f=2$) NJL results. It is important to mention that the NJL model has limitations concerning some phenomena, like confinement or inverse magnetic catalysis. And modifications have been made to the model in order to improve its applications. However, future improvements are expected in both model and lattice calculations, where in the later, one could explore ways to circumvent problems arising with mixing states, while in the former it is possible to explore the effect of a magnetic field dependent coupling and beyond mean field calculations. Both approaches are far beyond the scope of the present manuscript and will be left for future works.
 Besides, we have also obtained that the non-polarized $\rho^0$ mass increases as a function of the magnetic field mainly due to the  components with spin projection $s_z=\pm 1$  which are responsible for a $2/3$ fraction of its mass. 

 We also present some results  to clarify the role of the free parameter $G_v$
 in the calculation of the $\rho^0$ meson mass.
 To obtain a bound state solution for the three  projections considered,
 we have to be constrained by $m_{\rho^0}^2<4M^2$ and this relation is not satisfied for all  the range of magnetic fields of interest, i.e, depending on the value of the
 adopted $G_v$.  Of course, the choice of the $G_v$ values in order to guarantee that the $\rho$ mass is a bound state solution has to be considered with care, since the vacuum value ($eB=0$) of the physical $\rho^0$ mass also depends on this choice.
 Therefore, the parametrization is not fully free of inconsistency. However, given the full range of magnetic field available, and depending on the
 choice of original parameters, we think it is possible to handle  the free parameter $G_v$ in order to obtain a more reasonably vacuum values for the $\rho^0$
mass. In the near future, we expect to explore the role of confinement and finite temperature in the $\rho^0$ meson mass. 

\section*{Acknowledgments}

This work was partially supported by Conselho Nacional de Desenvolvimento Cient\'ifico 
e Tecno\-l\'o\-gico  (CNPq), Grants No. 309598/2020-6 (R.L.S.F.), No. 304518/2019-0 (S.S.A.), No. 306615/2018-5 (V.S.T.); Coordena\c c\~{a}o  de 
Aperfei\c coamento de Pessoal de  N\'{\i}vel Superior - (CAPES) Finance  Code  001 ( W.R.T); 
Funda\c{c}\~ao de Amparo \`a Pesquisa do Estado do Rio 
Grande do Sul (FAPERGS), Grants Nos. 19/2551- 0000690-0 and 19/2551-0001948-3 (R.L.S.F.);
Funda\c{c}\~ao de Amparo \`a Pesquisa do Estado de S\~ao Paulo (FAPESP), No. 2019/10889-1 (V.S.T.); Fundo de Apoio ao Ensino, Pesquisa e \`a Extens\~{a}o (FAEPEX), Grant No. 3258/19 (V.S.T.). 
The work is also part of the project Instituto Nacional de Ci\^encia 
e Tecnologia - F\'isica Nuclear e Aplica\c{c}\~oes (INCT - FNA), Grant No. 464898/2014-5.





\end{document}